\documentclass[aps,prl,twocolumn,groupedaddress,showpacs]{revtex4}
\usepackage{graphicx}
\usepackage{dcolumn}
\usepackage{bm}
\input epsf
\epsfclipon

\begin{document}
\title{Average path length in random networks}
\author{Agata Fronczak, Piotr Fronczak and Janusz A. Ho\l yst}
\affiliation{Faculty of Physics and Center of Excellence for
Complex Systems Research, Warsaw University of Technology,
Koszykowa 75, PL-00-662 Warsaw, Poland}
\date{\today}
\begin{abstract}
Analytic solution for the average path length in a large class of
random graphs is found. We apply the approach to classical random
graphs of Erd\"{o}s and R\'{e}nyi $(ER)$ and to scale-free
networks of Barab\'{a}si and Albert $(BA)$. In both cases our
results confirm previous observations: small world behavior in
classical random graphs $l_{ER} \sim \ln N$ and ultra small world
effect characterizing scale-free $BA$ networks $l_{BA} \sim \ln
N/\ln\ln N$. In the case of scale-free random graphs with power
law degree distributions we observed the saturation of the average
path length in the limit of $N\rightarrow\infty$ for systems with
the scaling exponent $2< \alpha <3$ and the small-world behaviour
for systems with $\alpha>3$.
\end{abstract}  \maketitle

During the last few years random, evolving networks have become a
very popular research domain among physicists \cite{0a,0b,1,2}. A
lot of efforts were put into investigation of such systems, in
order to recognize their structure and to analyze emerging complex
properties. It was observed that despite network diversity, most
of real web-like systems share three prominent features: small
average path length ($APL$), high clustering and skewed degree
distribution \cite{0a,0b,1,2,14}. Several network topology
generators have been proposed to embody the fundamental network
characteristics \cite{22,12,15,16}. Due to extensive numerical
simulations there were created and analyzed realistic models of
real networks especially basing on preferential attachment rule
introduced by Barab\'{a}si and Albert \cite{1,2,22}. The most
basic issues within the scope of network investigation are
structural: connectivity distributions \cite{22,16,11},
correlation analyzes (including clustering) \cite{5a,7,20} and
finally estimations of the $APL$ \cite{7a,8,9,9a}. The last
characteristics is of great importance for network studies as it
delivers basic information on a type of network geometry. It is
clear, that a better understanding of network topology is of great
importance for modern network designing and indirectly affects
such crucial fields like information processing in different
communication systems (including the Internet) \cite{26,27,28,30},
disease or rumor transmission in social networks \cite{33,34,35}
and network optimization \cite{29,31,46}. All these processes
become more efficient when the mean distance between network sites
is smaller.

It is well known that random networks such as Erd\"{o}s and
R\'{e}nyi ($ER$) graphs, as well as partially random networks such
as Watts-Strogatz {\it small-world} models \cite{32,7a}, have a
very small $APL$, which scales as $l\sim\ln N$, where $N$
describes the network size. In fact, it was expected that the
logarithmic size effect on the $APL$ is a common property of
random networks \cite{8}. Very recently Cohen and Havlin found
\cite{9} that random networks with power-law degree distribution
$P(k)\sim k^{-\alpha}$ and the scaling exponent $2<\alpha<3$
exhibit anomalous scaling of the average distance $l\sim\ln\ln N$.
Such an anomalous scaling is expected to lead to anomalies in
diffusion and transport phenomena within the networks. The result
is particularly interesting since it is known that most of real
networks, including both manmade communication networks like the
Internet and natural networks like food or metabolic networks,
exhibit scale-free character with the relevant scaling exponents
\cite{0a,0b,1,2}.

The paper presents an analytic theory describing metric features
of random networks. It allows to calculate the main network
characteristics like: $APL$, intervertex distance distribution and
the mean number of vertices at a certain distance away from a
randomly chosen vertex. We compare our analytic results with
numerical simulations performed for $ER$ random graphs and for
scale-free Barab\'{a}si and Albert ($BA$) networks.

Let us start with the following lemma.
\newtheorem{tw}{Lemma}
\begin{tw}\label{tw1}
If $A_{1},A_{2},\dots,A_{n}$ are mutually independent events and
their probabilities fulfill relations $\forall_{i} P(A_{i})\leq
\varepsilon$ then
\begin{equation}
P(\bigcup_{i=1}^{n}A_{i})=1-\exp(-\sum_{i=1}^{n}P(A_{i}))-Q,
\end{equation}
where $0\leq Q<\sum_{j=0}^{n+1}
(n\varepsilon)^{j}/j!-(1+\varepsilon)^{n}$.
\end{tw}
{\bf Proof.} Using the method of inclusion and exclusion \cite{38}
we get
\begin{eqnarray}\label{p1}
P(\bigcup_{i=1}^{n}A_{i})=\sum_{j=1}^{n}(-1)^{j+1}S(j),
\end{eqnarray}
with
\begin{eqnarray}\label{p2}
S(j)=\sum_{1 \leq i_{1} < i_{2} < \dots < i_{j} \leq n}^{n}
P(A_{i_{1}})P(A_{i_{2}})\dots P(A_{i_{j}}) \nonumber \\
=\frac{1}{j!}\left ( \sum_{i=1}^{n} P(A_{i}) \right )^{j}-Q_{j},
\end{eqnarray}
where $0\leq Q_{j}\leq \left( n^{j}/j! - \left( ^{n}_{j}
\right)\right)\varepsilon^{j}$. The term in bracket represents the total number
of redundant components occurring in the last line of (\ref{p2}). Neglecting
$Q_{j}$ it is easy to see that $(1-P(\cup A_{i}))$ corresponds to the first
$(n+1)$ terms in the MacLaurin expansion of $\exp(-\sum P(A_{i}))$. The effect
of higher-order terms in this expansion is smaller than
$R<(n\varepsilon)^{n+1}/(n+1)!$. It follows that the total error of (\ref{tw1})
may be estimated as $Q<\sum_{j=1}^{n}Q_{j}+R$. This completes the proof.

Let us notice that the terms $Q_{j}$ in (\ref{p2}) disappear when
one approximates multiple sums $\sum_{1 \leq i_{1} < i_{2} < \dots
< i_{j} \leq n}^{n}$ by corresponding multiple integrals. For
$\varepsilon = A/n\ll 1$ the error of the above assessment is less
then $A^{2}\exp(A)/n$ and may be dropped in the limit
$n\rightarrow \infty$.

A random graph with a given degree distribution $P(k)$ is the
simplest network model \cite{8}. In such a network the total
number of vertices $N$ is fixed. Degrees of all vertices are
independent identically distributed random integers drawn from a
specified distribution $P(k)$ and there are no vertex-vertex
correlations. Because of the lack of correlations the probability
that there exists a walk of length $x$ crossing index-linked
vertices $\{i,v_{1},v_{2} \dots v_{(x-1)},j\}$ is described by the
product
$\widetilde{p}_{iv_{1}}\;\widetilde{p}_{v_{1}v_{2}|iv_{1}}\;
\widetilde{p}_{v_{2}v_{3}|v_{1}v_{2}}\dots
\widetilde{p}_{v_{(x-1)}j|v_{(x-2)}v_{(x-1)}}$ where
\begin{equation}\label{pij1}
\widetilde{p}_{ij}=\frac{k_{i}k_{j}}{\langle k\rangle N},
\end{equation}
gives a connection probability  between vertices $i$ and $j$ with
degrees $k_{i}$ and $k_{j}$ respectively, whereas
\begin{equation}\label{pij1b}
\widetilde{p}_{ij|li}=\frac{(k_{i}-1)k_{j}}{\langle k\rangle N}
\end{equation}
describes the conditional probability of a link $\{i,j\}$ given
that there exists {\it another} link $\{l,i\}$. It is important to
stress that the graph theory distinguishes {\it a walk} from {\it
a path} \cite{45}. A walk is just a sequence of vertices. The only
condition for such a sequence is that two successive nodes are the
nearest neighbors. A walk is termed a path if all of its vertices
are distinct. In fact we are interested in the shortest paths. Let
us consider the situation when there exists at least one walk of
the length $x$ between the vertices $i$ and $j$. If the walk(s)
is(are) the shortest path(s) $i$ and $j$ are exactly $x$-th
neighbors otherwise they are closer neighbors. In terms of
statistical ensemble of random graphs \cite{5} the probability
$p_{ij}(x)$ of at least one walk of the length $x$ between $i$ and
$j$ expresses also the probability that these nodes are neighbors
of order not higher than $x$. Thus, the probability that $i$ and
$j$ are exactly $x$-th neighbors is given by the difference
\begin{equation}\label{pijx*A}
p_{ij}^{*}(x)=p_{ij}(x)-p_{ij}(x-1).
\end{equation}

In order to write the formula for $p_{ij}(x)$ we take advantage of
the lemma (\ref{tw1})
\begin{eqnarray}\label{pijxA}
p_{ij}(x)=1-Q-\hspace{5.5cm} \\ \nonumber
\exp[-\sum_{v_{1}=1}^{N}\sum_{v_{2}=1}^{N}\dots
\sum_{v_{(x-1)}=1}^{N}\widetilde{p}_{iv_{1}} \dots
\widetilde{p}_{v_{(x-1)}j|v_{(x-2)}v_{(x-1)}}],
\end{eqnarray}
where $N$ is the total number of vertices in a network. A sequence
of $(x+1)$ vertices $\{i,v_{1},v_{2}\dots ,v_{(x-1)},j\}$
beginning with $i$ and ending with $j$ corresponds to a single
event $A_{i}$ and the number of such events is given by
$n=N^{x-1}$. Putting (\ref{pij1}) into (\ref{pijxA}) and replacing
the summing over nodes indexes by the summing over the degree
distribution $P(k)$ one gets:
\begin{equation}\label{pijxB}
p_{ij}(x)=1-\exp\left[-\frac{k_{i}k_{j}}{N}\frac{\langle k(k-1)
\rangle^{x-1}}{\langle k\rangle^{x}}\right]-Q.
\end{equation}

The assumption underlying (\ref{tw1}) is the mutual independence
of all contributing events $A_{i}$. In fact, since the same edge
may participate in several $x-$walks there exist correlations
between these events. Nevertheless, it is easy to see that the
fraction of correlated walks is negligible for short walks ($x\ll
N$) that play the major role in random graphs showing small-world
behavior.

The question is when the term $Q$ in (\ref{pijxB}) may be neglected. To work out
the problem let us perform the following reasoning: if $\forall_{(i,j)}$ there
exists $\widetilde{\varepsilon}\ll 1$ such that $\widetilde{p}_{ij}\leq
\widetilde{\varepsilon}$ then $\forall_{x\geq 1}$
$\widetilde{p}_{iv_{1}}\;\widetilde{p}_{v_{1}v_{2}|iv_{1}}\dots
\widetilde{p}_{v_{(x-1)}j|v_{(x-2)}v_{(x-1)}}\leq
\widetilde{\varepsilon}^{\:x}\ll 1$ and $Q$ may be ignored. In fact, due to
(\ref{pij1}) the condition $\widetilde{p}_{ij}\ll 1$ is not fulfilled for pairs
of vertices $i$ and $j$ possessing large degrees $k_{i}$ and $k_{j}$. The
fraction of such pairs may be estimated as
\begin{equation}\label{rgcond}
\int_{k_{min}}^{k_{max}}P(k_{j})\int_{\widetilde{\varepsilon}\langle
k\rangle N/k_{j}}^{k_{max}} P(k_{i})dk_{i}dk_{j}\ll 1.
\end{equation}
Using the Chebyshev's inequality \cite{38} and solving
(\ref{rgcond}) with respect to $\widetilde{\varepsilon}\ll 1$ one
gets the condition when $Q$ may be dropped
\begin{equation}\label{cond}
\frac{\langle k^{2}\rangle}{\langle k\rangle^{2}}(\langle k^{2}
\rangle-\langle k\rangle^{2})\ll N^{2}.
\end{equation}

Due to (\ref{pijx*A}) the probability that both vertices are
exactly the $x$-th neighbors may be written as
\begin{equation}\label{pijx*B}
p_{ij}^{*}(x)=F(x-1)-F(x),
\end{equation}
where
\begin{equation}\label{Fx}
F(x)=\exp \left[ -\frac{k_{i}k_{j}}{N} \frac{(\langle
k^{2}\rangle-\langle k\rangle)^{x-1}}{\langle k\rangle^{x}}
\right].
\end{equation}
Note that averaging (\ref{pijx*B}) over all pairs of vertices one
may obtain the intervertex distance distribution
$p(x)=\langle\langle p_{ij}^{*}(x)\rangle_{i}\rangle_{j}$. Now the
mean number of vertices  at a certain distance $x$ away from a
randomly chosen vertex $i$ can be written as $z_{x}=\int
p_{ij}^{*}(x)P(k_{j})N dk_{j}$. Taking only the first two terms of
power series expansion of both exponential functions in
(\ref{pijx*B}) one gets the relationship obtained by Newman et al.
\cite{8} $z_{x}=(z_{2}/z_{1})^{x-1}z_{1}$ that was received
assuming a tree-like structure of random graphs.

The expectation value for the $APL$ between $i$ and $j$ is
\begin{equation}\label{lijA}
l_{ij}(k_{i},k_{j})=\sum_{x=1}^{\infty}\:x\:p_{ij}^{*}(x)=\sum_{x=0}^{\infty}F(x).
\end{equation}
Notice that a walk may cross the same node several times thus the
largest possible walk length can be $x=\infty$. The Poisson
summation formula allows us to simplify (\ref{lijA})
\begin{equation}\label{lij}
l_{ij}(k_{i},k_{j})=\frac{-\ln k_{i}k_{j}+\ln(\langle
k^{2}\rangle-\langle k\rangle)+\ln N -\gamma}{\ln(\langle
k^{2}\rangle/\langle k\rangle-1)}+\frac{1}{2},
\end{equation}
where $\gamma\simeq 0.5772$ is the Euler's constant. The average
intervertex distance for the whole network depends on a specified
degree distribution $P(k)$
\begin{equation}\label{lRG}
 l=\frac{\ln(\langle k^{2}\rangle-\langle k\rangle)-2\langle\ln k\rangle+\ln N
-\gamma}{\ln(\langle k^{2}\rangle/\langle
k\rangle-1)}+\frac{1}{2}.
\end{equation}
The formulas (\ref{lij}) and (\ref{lRG}) diverge when $\langle
k^{2}\rangle=2\langle k\rangle$, giving the well-known estimation
of percolation threshold in undirected random graphs
\cite{pc1,pc2}.

To test the formula (\ref{lRG}) we start with two well known
networks: $ER$ classical random graphs and scale-free $BA$
networks. The choice of these two networks is not accidental. Both
models play an important role in the network science
\cite{0a,0b,1,2}. The $ER$ model was historically the first one
but it has been realized it is too random to describe real
networks. The most striking discrepancy between $ER$ model and
real networks appears when comparing degree distributions. As
mention at the beginning of the paper degree distribution follows
power-law in most of real systems, whereas classical random graphs
exhibit Poisson degree distribution. The only known mechanism
driving real networks into scale-free structures is preferential
attachment. The simplest model that incorporates the rule of
preferential attachment was originally introduced by Barab\'{a}si
and Albert \cite{22}.

{\it Classical $ER$ random graphs}. For these networks the degree
distribution is given by the Poisson function $P(k)=e^{-\langle
k\rangle}\langle k\rangle^{k}/k!$ and the condition (\ref{cond})
is always fulfilled. However, since $\langle\ln k\rangle$ cannot
be calculated analytically for Poisson distribution thus the $APL$
may not be directly obtained from (\ref{lRG}). To overcome this
problem we take advantage of the mean field approximation. Let us
assume that all vertices within a graph possess the same degree
$\forall_{i}\:k_{i}=\langle k\rangle$. It implies that the $APL$
between two arbitrary nodes $i$ and $j$ (\ref{lRG}) should
describe the average intervertex distance of the whole network
\begin{equation}\label{lER}
l_{ER}=\frac{\ln N - \gamma}{\ln(pN)}+\frac{1}{2}.
\end{equation}

\begin{figure} \epsfxsize=6.4cm \epsfbox{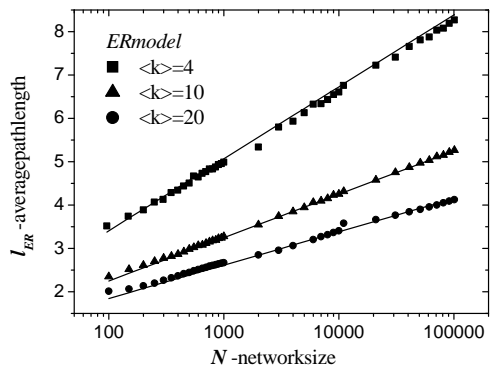}
\caption{The average path length $l_{ER}$ versus network size $N$
in $ER$ classical random graphs with $\langle k \rangle
=pN=4,10,20$. The solid curves represent numerical prediction of
Eq.(\ref{lER}).} \label{figer}
\end{figure}

Until now only a rough estimation of the quantity has been known.
One has expected that the average shortest path length of the
whole ER graph scales with the number of nodes in the same way as
the network diameter. We remind that the diameter $d$ of a graph
is defined as the maximal distance between any pair of vertices
and  $d_{ER}=\ln N/\ln(pN)$ \cite{1,2}. Fig.\ref{figer} shows the
prediction of the equation (\ref{lER}) in comparison with the
numerically calculated $APL$ in classical random graphs.

{\it Scale-free $BA$ networks}. The basis of the $BA$ model is its
construction procedure. Two important ingredients of the procedure
are: continuous network growth and preferential attachment. The
network starts to grow from an initial cluster of $m$ fully
connected vertices. Each new node that is added to the network
creates $m$ links that connect it to previously added nodes. The
preferential attachment means that the probability of a new link
growing out of a vertex $i$ and ending up in a vertex $j$ is given
by $\widetilde{p}_{ij}^{BA}=mk_{j}(t_{i})/\sum_{l}k_{l}(t_{i})$,
where $k_{j}(t_{i})$ \cite{42} denotes the connectivity of a node
$j$ at the time when a new node $i$ is added to the network.
Taking into account the time evolution of node degrees in $BA$
networks one can show  that the probability
$\widetilde{p}_{ij}^{BA}$ is equivalent to (\ref{pij1}). Now let
us consider the conditional probability $\widetilde{p}_{ij|li}$.
Checking the possible time order of the vertices $i,j,l$ it is
easy to see that in five of $3!$ cases
$\widetilde{p}_{ij|li}=\widetilde{p}_{ij}$ and in a good
approximation we get instead of  (\ref{pijxB}) the result
\begin{equation}\label{pijxBA}
p_{ij}^{BA}(x)=1-\exp \left[ -\frac{k_{i}k_{j}}{N} \frac{\langle
k^{2}\rangle^{x-1}}{\langle k\rangle^{x}} \right].
\end{equation}

\begin{figure} \epsfxsize=6.6 cm \epsfbox{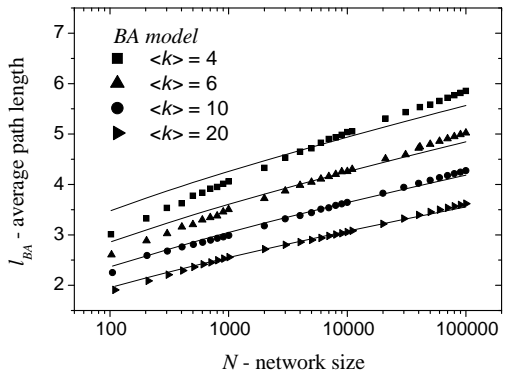}
\caption{Characteristic path length $l_{BA}$ versus network size
$N$ in $BA$ networks. Solid lines represent Eq.(\ref{lBA}).}
\label{figba}
\end{figure}

It was found \cite{42} that the degree distribution in $BA$
network is given by $P(k)=2m^{2}k^{-\alpha}$, where
$k=m,m+1,\dots,m\sqrt{N}$, and the scaling exponent $\alpha=3$.
Putting $\langle k\rangle=2m$, $\langle k^{2}\rangle=m^{2}\ln N$
and taking into account (\ref{pijxBA}) one gets that the $APL$
between $i$ and $j$ is given by
\begin{equation}\label{lijBA}
l_{ij}^{BA}(k_{i},k_{j})=\frac{-\ln(k_{i}k_{j})+\ln N+\ln(2m)
-\gamma }{\ln\ln N+\ln (m/2)}+\frac{3}{2}.
\end{equation}
Averaging (\ref{lijBA}) over all vertices we obtain
\begin{equation}\label{lBA}
l_{BA}=\frac{\ln N-\ln(m/2)-1-\gamma}{\ln\ln
N+\ln(m/2)}+\frac{3}{2}.
\end{equation}
Fig.\ref{figba} shows the $APL$ of $BA$ networks as a function of
the network size $N$ compared with the analytical formula
(\ref{lBA}). There is a visible discrepancy between the theory and
numerical results when $\langle k\rangle=4$. The discrepancy
disappears when the network becomes denser i.e. when $\langle
k\rangle$ increases. We suspect that it is the effect of
structural correlations that occur within the evolving networks
\cite{5a,x1} and are absent in random graphs used for analytic
calculations. The results let us deduce that the correlations
become less important in denser networks.

{\it Scale-free networks with arbitrary scaling exponent}. Let us
consider scale-free random graphs with degree distribution given
by a power law, i.e.
$P_\alpha(k)=(\alpha-1)m^{\alpha-1}k^{-\alpha}$, where
$k=m,m+1,\dots,mN^{1/(\alpha-1)}$ \cite{9}. Solving (\ref{cond})
for $P_\alpha(k)$ one can see  that our approach should work for
$\alpha>2$. Taking advantage of (\ref{lRG}) we get that for large
networks $N\gg 1$ the $APL$ scales as follows
\begin{itemize}
\item $l\simeq 2/(3-\alpha)+1/2\:\:\:$ for $2<\alpha<3$,
\item $l\simeq \ln N/\ln\ln N+3/2\:\:\:$ for $\alpha=3$,
\item $l\simeq \ln N/(\ln(m(\alpha-2)/(\alpha-3)-1)+1/2\:\:\:$ for $\alpha>3$.
\end{itemize}
The result for $\alpha\geq 3$ is consistent with estimations
obtained by Cohen and Havlin \cite{9}. The first case with $l$
independent on $N$ shows that there is a  saturation effect for
the mean path length in large networks. Note, that the effect was
observed in metabolic networks \cite{x}.

In conclusion, we presented a theory for metric properties of
random networks with arbitrary degree distribution. The approach
is applied to get an analytic formula for the $APL$ in a large
class of undirected random graphs with an arbitrary degree
distribution $P(k)$. The results are in a very good agreement with
numerical simulations performed for $ER$ random graphs and for
$BA$ networks. We observed saturation of $l$ in the limit
$N\rightarrow\infty$ of scale-free networks with scaling exponents
from the range $2<\alpha<3$,  the small-world behaviour for
networks with $\alpha>3$ and the ultra small-world behaviour of
$BA$ model. Our derivations show that the behaviour of $APL$
within scale-free networks is even more intriguring than reported
in the recent paper of Cohen and Havlin \cite{9}.

Appendix. After finishing the paper we  learned about the preprint
on this subject written by Dorogovtsev, Mendes and Samukhin
\cite{metric}. Basing on generating function formalism the authors
derived a similar formula for the $APL$ in random graphs $l\sim
\ln N/\ln(\langle k^{2}\rangle/\langle k\rangle-1)$.
\par Acknowledgments. We are thankful to Sergei Dorogovtsev for
critical comments to the preliminary version of this paper. One of
us (AF) thanks The State Committee for Scientific Research in
Poland for support under grant No. $2 P03B 013 23$.


\end{document}